\documentclass[prl,aps,amssymb,amsfonts,amsmath,showpacs,superscriptaddress,twocolumn]{revtex4}
\usepackage{graphicx, epstopdf} % Einbindung von Grafikdateien
\usepackage{amsfonts}%%% math fonts needed

\usepackage{graphicx,xcolor}
\usepackage{amsmath, bbm}
\usepackage{natbib}
\usepackage{braket}
\usepackage[utf8x]{inputenc}
\usepackage[ngerman, english]{babel}
\usepackage[T1]{fontenc}

\usepackage{hyperref}
\usepackage{subscript}

\newcommand{\mplaffil}{Max Planck Institute for the Science of Light (MPL), D-91058 Erlangen, Germany}
\newcommand{\HZBaffil}{Helmholtz Centre Berlin for Materials and Energy, D-14109 Berlin, Germany}
\newcommand{\fauaffil}{Department of Physics, Friedrich Alexander University of Erlangen-N\"urnberg, D-91058 Erlangen, Germany}

\begin{document} \title{A Sub-$\rm \lambda^{3}$-Volume Cantilever-based Fabry-P\'erot Cavity}
\author{Hrishikesh Kelkar}
\affiliation{\mplaffil}
\author{Daqing Wang}
\affiliation{\mplaffil}
\author{Diego Mart\'in-Cano}
\affiliation{\mplaffil}
\author{Bj\"orn Hoffmann}
\affiliation{\mplaffil}
\author{Silke Christiansen}
\affiliation{\mplaffil}
\affiliation{\HZBaffil}
\author{Stephan G\"otzinger}
\affiliation{\fauaffil}
\affiliation{\mplaffil}
\author{Vahid Sandoghdar}
\affiliation{\mplaffil}
\affiliation{\fauaffil}

\begin{abstract} We report on the realization of an open plane-concave Fabry-P\'erot resonator with a mode volume below $\lambda^3$ at optical frequencies. We discuss some of the less common features of this new microcavity regime and show that the ultrasmall mode volume allows us to detect cavity resonance shifts induced by single nanoparticles even at quality factors as low as $100$. Being based on low-reflectivity micromirrors fabricated on a silicon cantilever, our experimental arrangement provides broadband operation, tunability of the cavity resonance, lateral scanning and promise for optomechanical studies.
\end{abstract} 

\maketitle
\section{Introduction}

Starting with the question of whether the excited-state lifetime of an atom can be modified, physicists have continuously explored ways to engineer the radiative properties of quantum emitters. The first experimental studies were performed in the near field of flat interfaces \cite{drexhage68}, but the change of spontaneous emission is commonly associated with Purcell's prediction that a cavity of quality factor $Q$ and mode volume $V$ can accelerate the radiation of a dipolar transition by $F_{\rm p}=(3\lambda^3/4\pi^2)(Q/V)$ folds, where $\lambda$ is the wavelength in the corresponding medium \cite{Purcell:46}. Following this recipe, many clever resonator schemes such as high-Q open Fabry-P\'erot cavities (FPC), monolithic FPCs in form of pillars, whispering gallery mode resonators and photonic crystal structures have been investigated for realizing large Purcell factors \cite{Vahala-book}. Nevertheless, sizable modifications of the spontaneous emission process remain nontrivial because no cavity design has succeeded in providing all the decisive ingredients of large $Q$, small $V$ (ideally down to its fundamental value of the order of $(\lambda/2)^3$), a facile way of tuning the cavity resonance, and compatibility with emitters of different materials. Of the various cavity geometries, open FPCs remain particularly attractive because they nicely lend themselves to the latter two criteria. 

\begin{figure}[h]
\centering
\includegraphics[width=8.6cm]{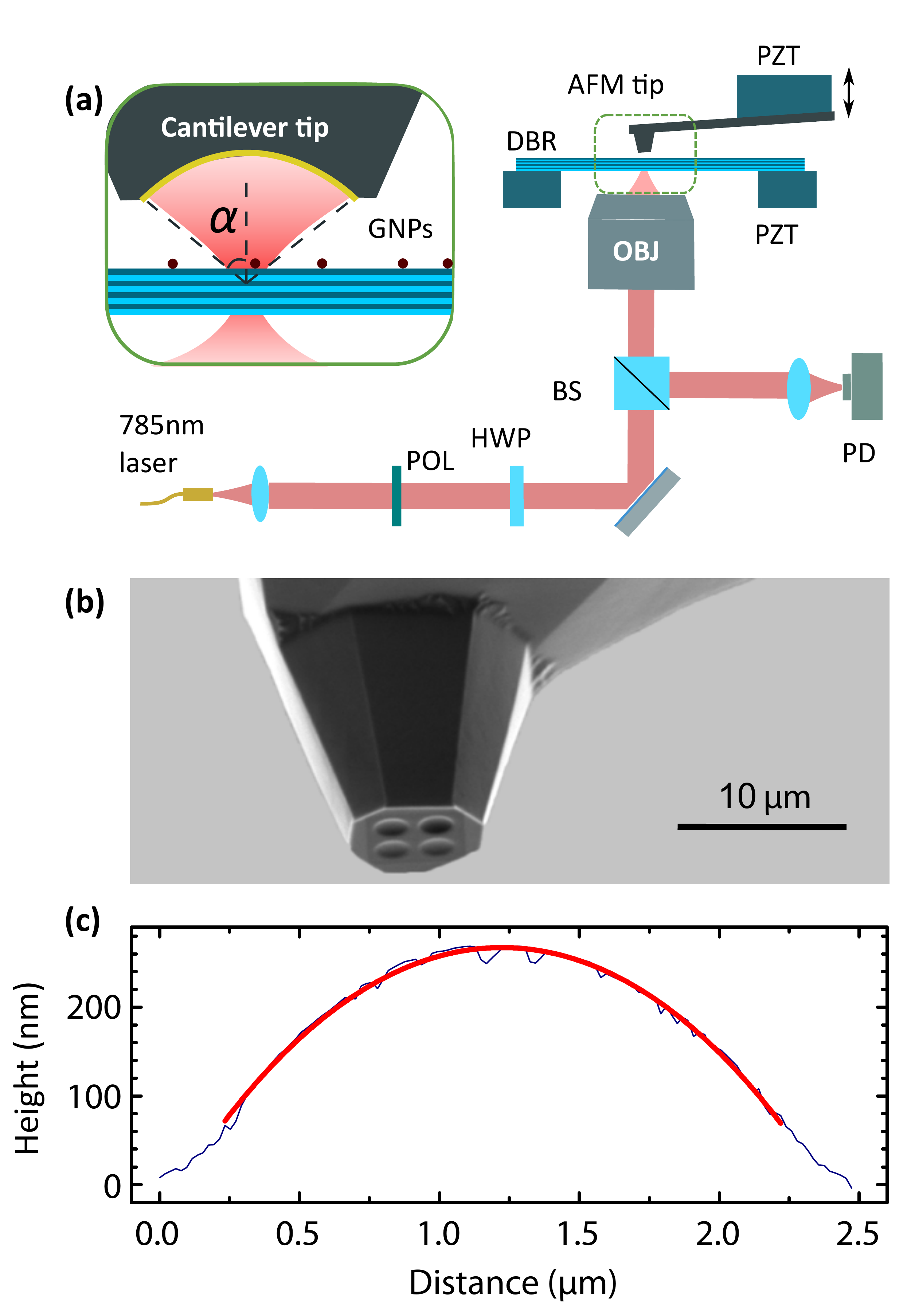}
\caption{(a) Schematics of the optical setup. OBJ: microscope objective; BS: beam splitter; HWP: half-wave plate; POL: linear polarizer; PD: photodiode; GNP: gold nanoparticle. The inset shows a zoom of the cavity region. (b) Scanning electron microscope image of four concave mirrors fabricated on a silicon pedestal at the end of a cantilever. (c) Atomic force microscope (AFM) cross section of a micromirror.}\label{setup}
\end{figure}

Efforts using open FPCs have usually pursued large $F_{\rm p}$ via high $Q$s instead of low $V$, but this choice is accompanied by several disadvantages. First, the resulting narrow cavity linewidths make the cavity extremely sensitive to mechanically, thermally or stress-induced length changes. Second, broad transitions in the condensed phase cannot couple to very narrow cavity resonances in an efficient manner. This is especially a restriction for room-temperature studies, where homogeneous linewidths of quantum dots, molecules or ions can reach several nanometers. In addition, scattering from finite-sized particles, e.g. from diamond nanocrystals, could quickly spoil the high Q of the cavity. Furthermore, narrow cavity resonances  do not allow simultaneous coupling to different transitions, for instance, for studies of $\Lambda$ transitions or of nonlinear interactions involving several light beams~\cite{Gruneisen:89}. 

In this article, we explore a new cavity design using a micromirror with a radius of curvature as small as 2.6~$\mu \rm m$ fabricated on a cantilever. By realizing a microcavity with a sub-$\lambda^3$ volume and low $Q$, we obtain a sizable $F_{\rm p}$ that would lead to the modification of the spontaneous emission rate by more than one order of magnitude while keeping a spectral bandwidth as broad as 1 THz. The second novel feature of our arrangement is the compatibility of its high numerical aperture ($\rm NA$) with the recent advances in efficient atom-photon coupling via tight focusing~\cite{Wrigge:08,Zumofen:08,Tey:08}.  Third, the open character, scanning capability and low Q of our FPC provide the possibility of using it as a scanning microscope~\cite{Toninelli:10b, Mader:14}, which can be used for local field-enhanced spectroscopy~\cite{hartschuh08} and sensing~\cite{Vollmer:08,Oezdemir:14, Mirsadeghi:14} of very different materials such as semiconductor quantum dots, carbon nanotubes, organic molecules, rare earth ions, and biomolecules. In particular, the low Q of the cavity ensures that even broad resonances of these materials couple efficiently to the cavity modes at room temperature. Fourth, the cantilever-based feature of our experimental arrangement makes it highly interesting for an unexplored regime of optomechanical investigations~\cite{Aspelmeyer:14} as we discuss in the concluding paragraph of this article.

\section{Key features of the Microcavity}

The inset in Fig.~\ref{setup}(a) shows the schematics of our cavity made of a flat distributed Bragg reflector (DBR) and a micromirror fabricated by focused ion beam milling. Here, we started with an n-doped silicon cantilever that contained a pedestal (diameter 8 $\mu \rm m$). The area of the micromirror to be milled was divided into pixels of diameter 5 nm. By controlling the ion dose for each pixel, we obtained a spherical surface profile, which was then polished in a final step. Figure~\ref{setup}(b) displays an electron microscope image of a cantilever pedestal with four mirrors of different radii of curvature, and Fig.~\ref{setup}(c) presents an exemplary topography cross section recorded with an AFM.  In this case, the mirror has a radius of curvature of $R=2.6~\mu \rm m$ accompanied by a root-mean-square surface roughness below 5 nm.  An opening aperture diameter of 2.4 $\mu \rm m$ provides a concave mirror with numerical aperture $\rm NA=0.68$ following the standard defintion of  $\rm NA=\rm \sin \theta$, where $\theta$ is the angle subtended by half the mirror aperture and $R/2$ is the mirror focal length. We note that fabrication of micromirrors has been previously reported for wet etching~\cite{Trupke:05}, laser ablation~\cite{Steinmetz:06, Toninelli:10b, Kaupp:13} and focused ion beam milling~\cite{Dolan:10, Di:12}, but the majority of the existing works report radii of curvature above ten microns, limiting both $\rm NA$ and $V$. To our knowledge, we provide the highest numerical aperture for a tunable microcavity reported to date.

The micromirrors were coated with 150 nm of gold followed by 50 nm of silicon dioxide as a protective layer, yielding a nominal reflectivity of 96\% at $\lambda=785$ nm. Taking into account the slight roughness of this mirror (see Figure~\ref{setup}(c)), we expect the reflectivity to be reduced to 95\% due to residual scattering~\cite{Bennett:61}. The DBR consisted of 11 layers of $\rm TiO_2/SiO_2$ stacks finished by a 22 nm layer of $\rm TiO_2$ to place the field maximum at the mirror surface. The resulting structure had a total thickness of 2.14 $\mu \rm m$ with its design band center at 710 nm and a band edge at 841 nm, leading to reflectivities of 99.9\% at $\lambda=785$ nm and 99.99\% at 745 nm. As displayed in Fig.~\ref{setup}(a), the microcavity was assembled with a piezoelectric transducer (PZT) stack for the axial displacement of the cantilever and a PZT scanner for the lateral positioning of the DBR. We point out that the asymmetry of the mirror reflectivities causes a suboptimal impedance matching, which can be easily alleviated by using lower reflectivity DBR. In this work, however, we have not been concerned with coupling efficiencies.

Intuitively speaking, we want to operate the cavity close to the condition, where the rays from a strongly focused incident beam are retroreflected by the curved micromirror. It follows that a useful figure of merit for mode matching becomes the opening arc half-angle $\alpha$ subtended by the mirror aperture (see the inset of Fig.~\ref{setup}(a)). We, thus, define $\rm NA_{\rm eff}=\rm sin \alpha$ as an effective numerical aperture, which amounts to 0.46 in our experiment. We note in passing that according to common textbook knowledge, the stability of a plane-concave FPC becomes compromised as the cavity length becomes comparable to and larger than the mirror curvature~\cite{Siegman-book}. 

\subsection{Cavity spectrum}
To perform spectroscopy on the microcavity, we focused laser beams at $\lambda$={785} nm and $\lambda$=745 nm through a microscope objective with $\rm NA=0.75$, whereby the waist of the incident laser beam was adjusted to match $\rm NA_{\rm eff}$. We then scanned the axial position of the cantilever and monitored the reflected light on a photodiode. 

\begin{figure}[h]
\centering
\includegraphics[width=8.8 cm]{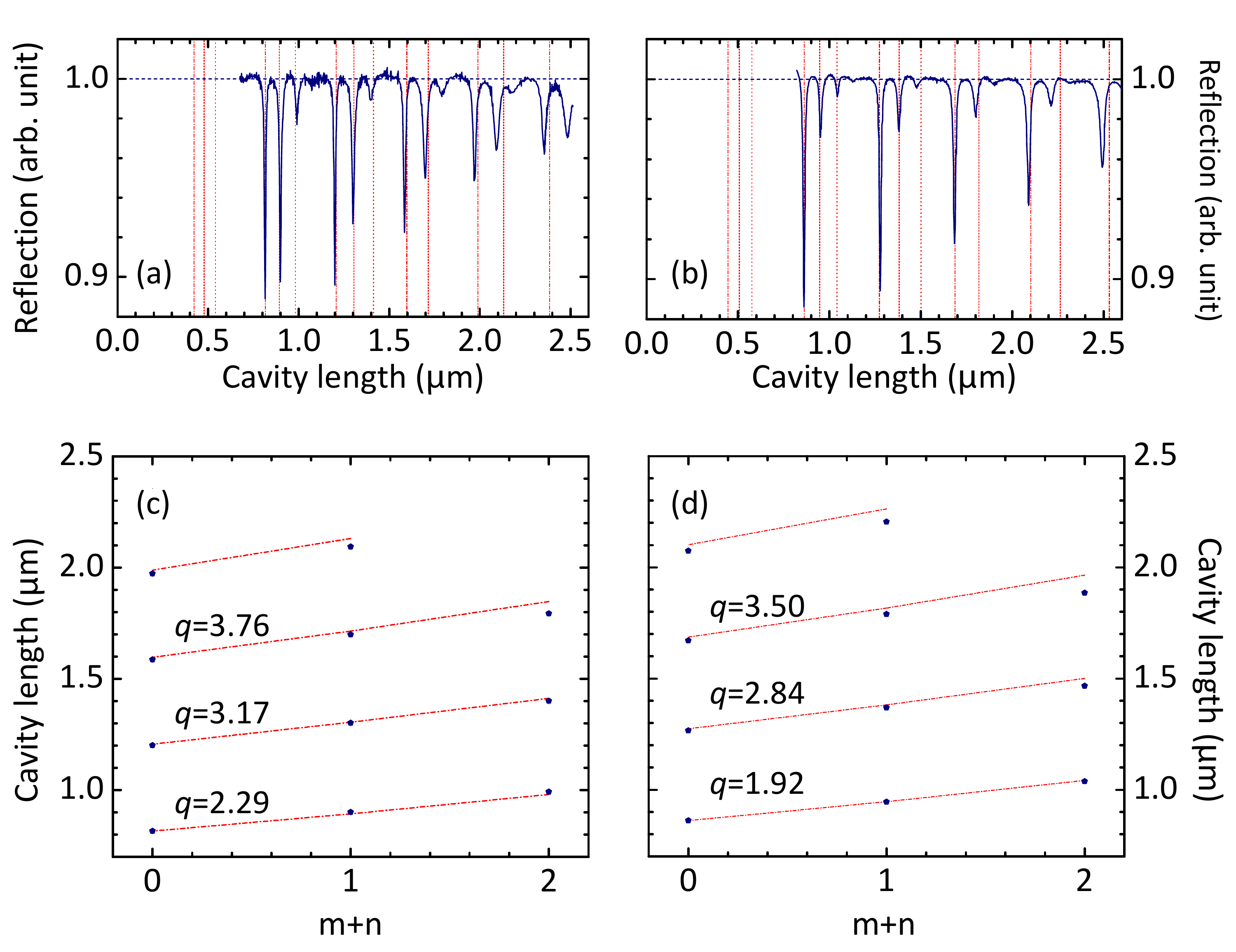}
\caption{Intensity of the light reflected from the cavity as a function of the cavity length change for wavelengths 745 nm (a) and 785 nm (b). The focus and the alignment of the incident beam were adjusted to ensure coupling to the higher-order transverse modes. The red vertical lines denote the resonances of the cavity according to a simple Hermite-Gaussian model. (c, d) Summary of the experimental and theoretical data from (a,b) plotted as a function of $\rm m+n$ for various $\rm TEM_{mn}$ modes. The indices $q$ signify the best fit obtained for the longitudinal mode number. }\label{cavity-spectra}
\end{figure}

The blue spectra in Figs.~\ref{cavity-spectra}(a) and (b) plot the measured reflection from the microcavity as a function of the change in $L$ at 745 nm and 785 nm, respectively. Here, we slightly defocused and misaligned the incident beam to obtain some coupling to the higher transverse modes. The red lines mark the resonances associated with the Hermite-Gaussian modes predicted by the equation~\cite{Svelto-book}
\begin{equation}
L=\frac{\lambda}{2}\Bigg[q+\frac{m+n+1}{\pi}\arccos\Bigg(\sqrt{1-\frac{L}{R}}\Bigg)\Bigg]~,
\label{CavityEqn1}
\end{equation}  
for a plane-concave cavity with the same nominal geometrical parameters as in the experiment but with infinitely thin mirrors. Here, the the term $\arccos{\sqrt{1-L/R}}$ signifies the Gouy phase within the paraxial approximation. The symbols in Fig.~\ref{cavity-spectra}(c) and (d) summarize the dependence of $L$ on $m+n$ indices of the transverse $\rm TEM_{mn}$ modes for each longitudinal parameter $q$. If we now fit these data with the outcome of Eq.~(\ref{CavityEqn1}), we obtain values for $q$ posted in the legends of Fig.~\ref{cavity-spectra}(c, d). The good cumulative comparison between theory and experiment suggests that we have reached the longitudinal mode $q=2$. 

\begin{figure}[h]
\centering
\includegraphics[width=8.5cm]{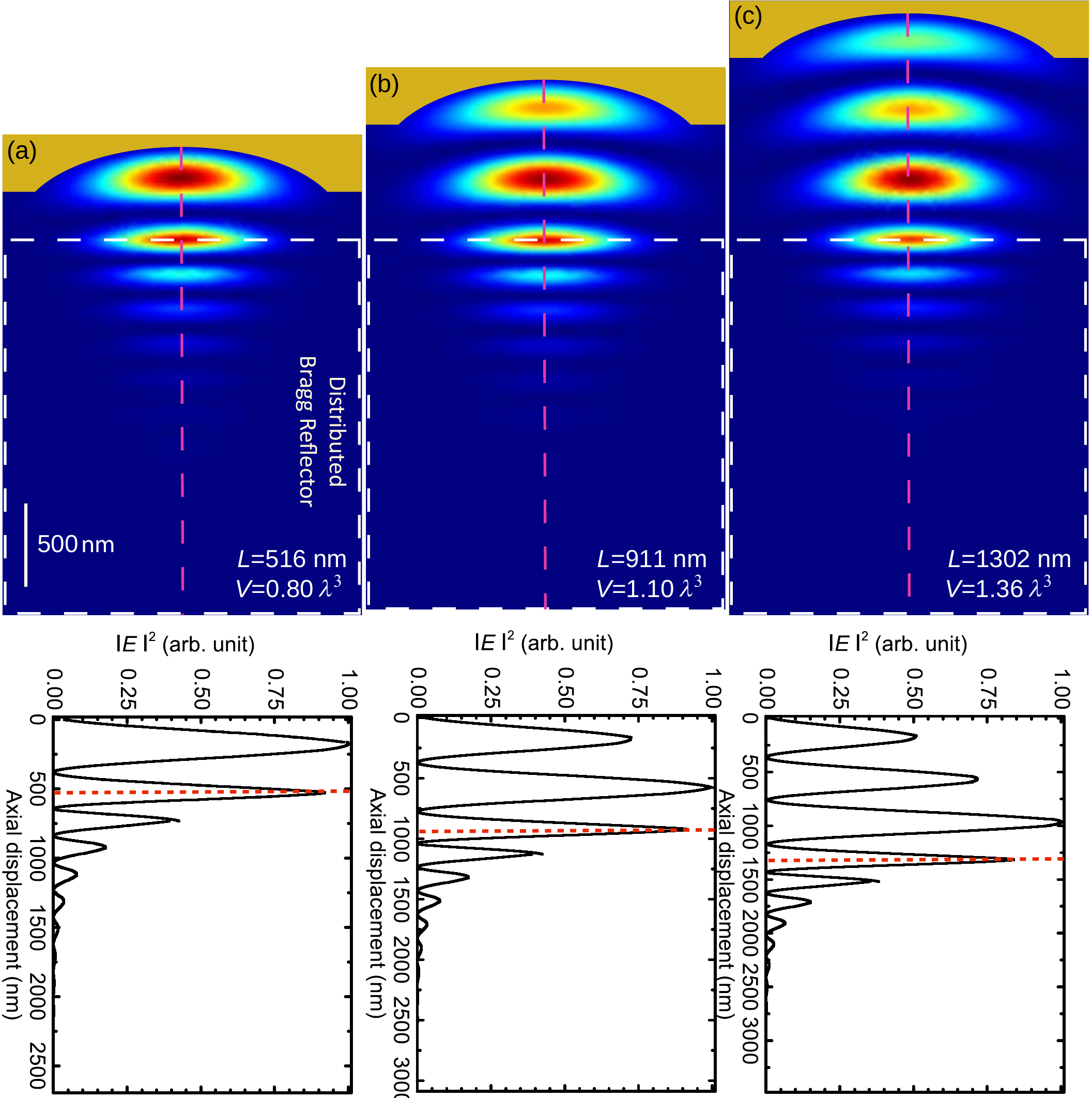}
\caption{Numerical simulation of the intensity for three different longitudinal modes $q=2, 3, 4$ at the wavelength of 745 nm. The cavity is formed by a gold-coated micromirror and a flat distributed Bragg reflector marked by the dashed lines. The length of the cavity $L$ measured between the air-DBR interface and the apex of the curved mirror is shown in each case together with a cut throught the intensity distribution along the red vertical dashed line. The red horizontal dotted lines mark the DBR interface.}
\label{volume-cuts}
\end{figure}

The resulting non-integer values of $q$ stem from the fact that Eq.~(\ref{CavityEqn1}) neglects the penetration depth of the field in the DBR mirror~\cite{Hood:01,Greuter:14} as illustrated in Fig.~\ref{volume-cuts} for $\lambda$= 745\,nm. The field penetration in the mirror and a considerable Gouy phase of $\pi/10$ in our high-NA cavity make the analysis of the spectra presented in Fig.~\ref{cavity-spectra} nontrivial. To account for these, we now examine the resonance condition and spectral properties of our cavity starting from basic derivations. 

The transmission spectrum of a Fabry-P\'erot cavity in the absence of absorption is given by the Airy function \cite{Silfvast:04}
\begin{equation}
	I_t=\frac{1}{1+F^\prime\sin^2(\Phi/2)}~,
	\label{eq:Airy}
\end{equation}
where $\Phi$ is the total round trip phase experienced by the light given by
\begin{equation}
	\Phi=2k\cdot L-\phi_\text{1}-\phi_{2}+2\phi_G~.
	\label{eq:RoundTripPhase}
\end{equation}
Here, $k=2\pi/\lambda$ is the wavevector, $\phi_{1}$ and $\phi_{2}$ are the phase shifts introduced by the two cavity mirrors, $L$ is the physical separation between the mirrors, and $\phi_G$ is the Gouy phase shift experienced by the mode after propagating the length $L$~\cite{footnote-Kelkar}. The cavity resonance occurs when $\Phi$ is an integer $q$ multiple of $2\pi$ such that sin($\Phi/2$)=0. At this point, one can define the finesse $\mathcal{F}$ as the ratio $\Delta \Phi/\delta \Phi$, where $\Delta \Phi=2\pi$ is the phase separation of two adjacent resonances, and $\delta \Phi$ is the full width at half-maximum (FWHM) of the phase through a given resonance. If this definition is applied to Eq.~(\ref{eq:Airy}), one obtains
\begin{equation}
\mathcal{F}=\pi\sqrt{F^{\prime}}/2=\frac{\pi\sqrt{\mathcal{R}}}{1-\mathcal{R}}~,
\label{eq:finesse}
\end{equation} 
where $\mathcal{R}$ is the geometric mean of the two mirror reflectivities. 

Equation~(\ref{eq:Airy}) provides an expression for the spectrum of the cavity. However, in the laboratory one does not have an easy  access to the phase of the light field. Instead, one either scans the frequency $\nu$ of the incident beam at a fixed $L$ or changes the latter at a fixed frequency. In most textbook treatments of Fabry-P\'erots, thin planar mirrors are considered, and the dependence of $\phi_G$, $\phi_\text{1}$ and $\phi_{2}$ on $L$ and $\nu$ is neglected. In that case, since only the propagation phase $kL$ changes, the spectra can be plotted as a function of $L$ or $\nu$ in a fully equivalent fashion, leading to $\mathcal{F}=\rm FSR(\nu)/\delta \nu=\rm FSR(L)/\delta L$. Here, $\rm FSR(L)$ and $\rm FSR(\nu)$ denote the separation of two adjacent resonance lengths and frequencies (commonly known as the free spectral range), respectively, whereas $\delta L$ and $\delta \nu$ signify the FWHM of the corresponding resonance as the cavity length or frequency are varied. This relation then allows a direct conversion of data obtained from cavity length spectroscopy to the properties of the frequency spectrum. 

\begin{figure}[!ht]
  \centerline{\includegraphics[width=88 mm]{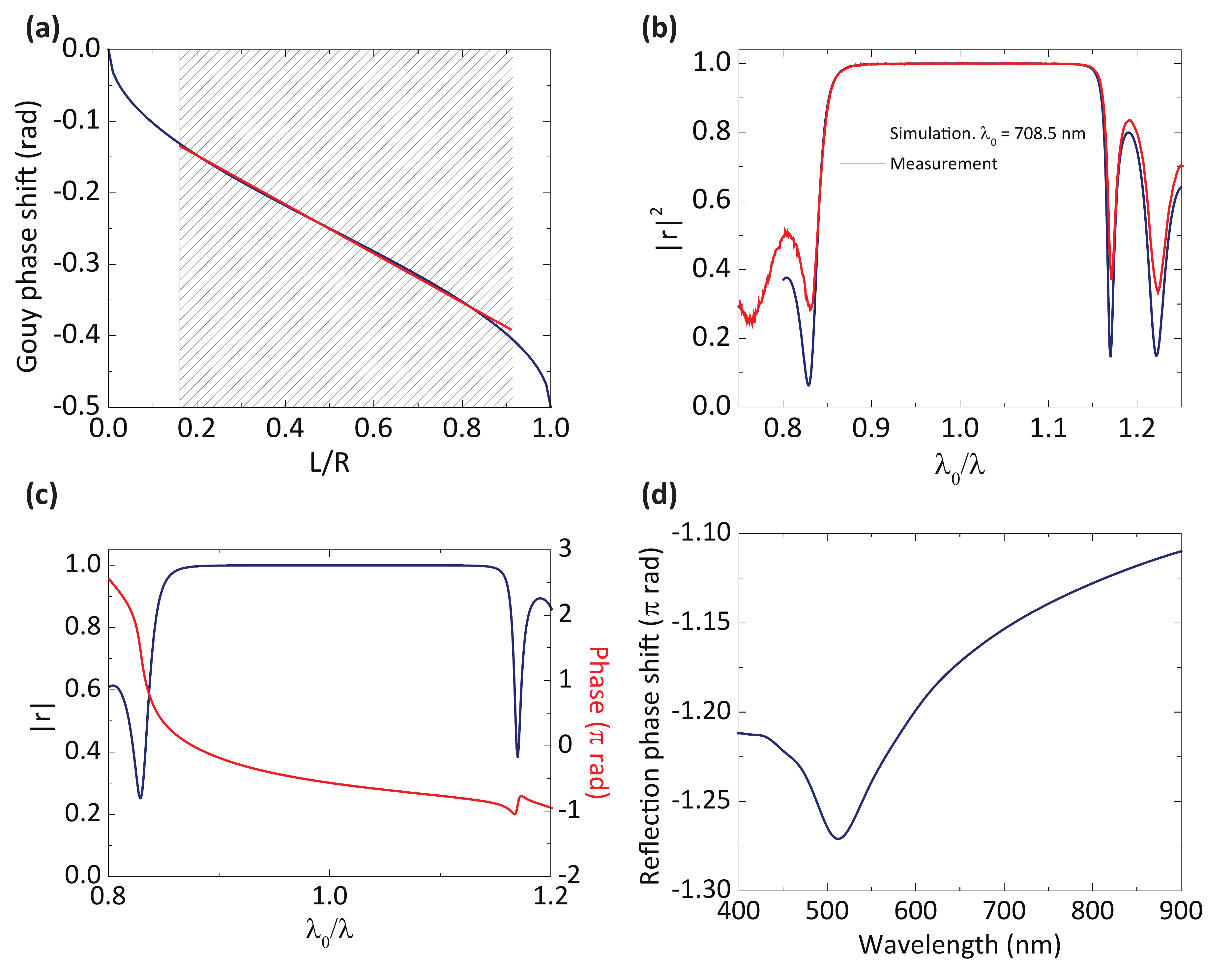}}
   \caption{(a) Gouy phase shift against the cavity length normalized to the radius of curvature of the mirror. Red line is a linear fit in the window of interest. (b) Measured reflection spectrum of the DBR (red) compared to a simulation (black) for a nominal design wavelength of $\lambda_0$ = 710 nm. The fit requires $\lambda_0$ = 708.5 nm. This slight discrepancy is likely due to finite tolerances in the DBR fabrication process. (c) Amplitude and phase of the complex reflectivity of the DBR. (d) Phase shift upon reflection from the gold mimrror.}
\label{Phases}
\end{figure}

We now allow $\phi_G$, $\phi_\text{1}$ and $\phi_{2}$ to depend on $L$ and $\nu$. It turns out that the dependence of $\phi_\text{1}$ and $\phi_{2}$ on $L$ is fairly negligible (as can be inferred from Fig.~\ref{volume-cuts}) and $\phi_{\rm G}$ varies linearly with $L$ in the regime of our cavity parameters, as shown in Fig. \ref{Phases}(a). In this case, Eq.~(\ref{eq:RoundTripPhase}) reads $\Phi \simeq 2k\cdot L-\phi_\text{1}-\phi_{2}+2(a+b L)=2k_{\rm eff}\cdot L-\phi_\text{1}-\phi_{2}+2a$, where $a$ and $b$ denote the zeroeth and first order terms of the Taylor expansion of $\phi_{\rm G}$, and $k_{\rm eff}$ signifies an effective frequency offset. In this case, the finesse can still be calculated as $\mathcal{F}=\rm FSR(L)/\delta L$, which is conveniently accessible in the experiment.

Next, we have to relate $\delta\nu$ and $\delta L$ in order to extract the quality factor $Q=\nu/\delta\nu$ from our experimental measurements.  We, thus, differentiate Eq. \ref{eq:RoundTripPhase} with respect to $\lambda$ at the cavity resonance to obtain
\begin{equation}
\frac{dL}{d\lambda}=\frac{1}{2}\left[\frac{2L}{\lambda}+\frac{\lambda}{2\pi}\left(\frac{d\phi_1}{d\lambda}+\frac{d\phi_2}{d\lambda}-2\frac{d\phi_G}{d\lambda}\right)\right]~.
\label{eq:LengthDispersion}
\end{equation}
The first term in the main paranthesis is the axial mode number $q$ of an ideal cavity, while the inner parenthesis contains corrections to it. In general, these corrections need not be small and can have implicit dependence on $\lambda$. Considering $\phi_G=a+bL$, we arrive at
\begin{equation}
Q=\left[\frac{2L}{\lambda}+\frac{\lambda}{2\pi}\left(\frac{d\phi_1}{d\lambda}+\frac{d\phi_2}{d\lambda}\right)\right] \mathcal{F}=q_{\rm eff}\cdot\mathcal{F}~
\label{eq:Q-F}
\end{equation} 
with~\cite{Babic:92}
\begin{equation}
q_{\rm eff}=\frac{2L}{\lambda}+\frac{\lambda}{2\pi}\left(\frac{d\phi_1}{d\lambda}+\frac{d\phi_2}{d\lambda}\right).
\label{eq:qeffexplicit}
\end{equation} 

To obtain the quantities in Eq.~(\ref{eq:qeffexplicit}), we measured the reflectivity spectrum of the DBR and compared the outcome with the calculations obtained from a transfer matrix method (see Fig.~\ref{Phases}(b)). The corresponding amplitude and phase of the reflection coefficient are shown in Fig.~\ref{Phases}(c). The phase shift upon reflection from the gold mirror is also calculated using the dielectric function from Ref.~\citep{johnson72} and is plotted in Fig. \ref{Phases}(d). The phase properties of the mirrors at the working wavelengths are summarized in Table \ref{table:MirrorPhases}.

\begin{table}[!ht]
\small\caption{Phase properties of the mirrors}
\vspace{0.1cm}
\centering
\begin{tabular}
{|c| c |c|}\hline 
Wavelength & 745 nm & 785 nm \\\hline %[0.5ex] % inserts table %heading\hline\hline
$\phi_\text{DBR}$& -0.5661$\pi$ & -0.1960$\pi$ \\\hline
$\phi_\text{metal}$& -1.1405$\pi$& -1.1310$\pi$\\\hline
\end{tabular}
\label{table:MirrorPhases}
\end{table}

\begin{figure}[h]
\centering
\includegraphics[width=8.0 cm]{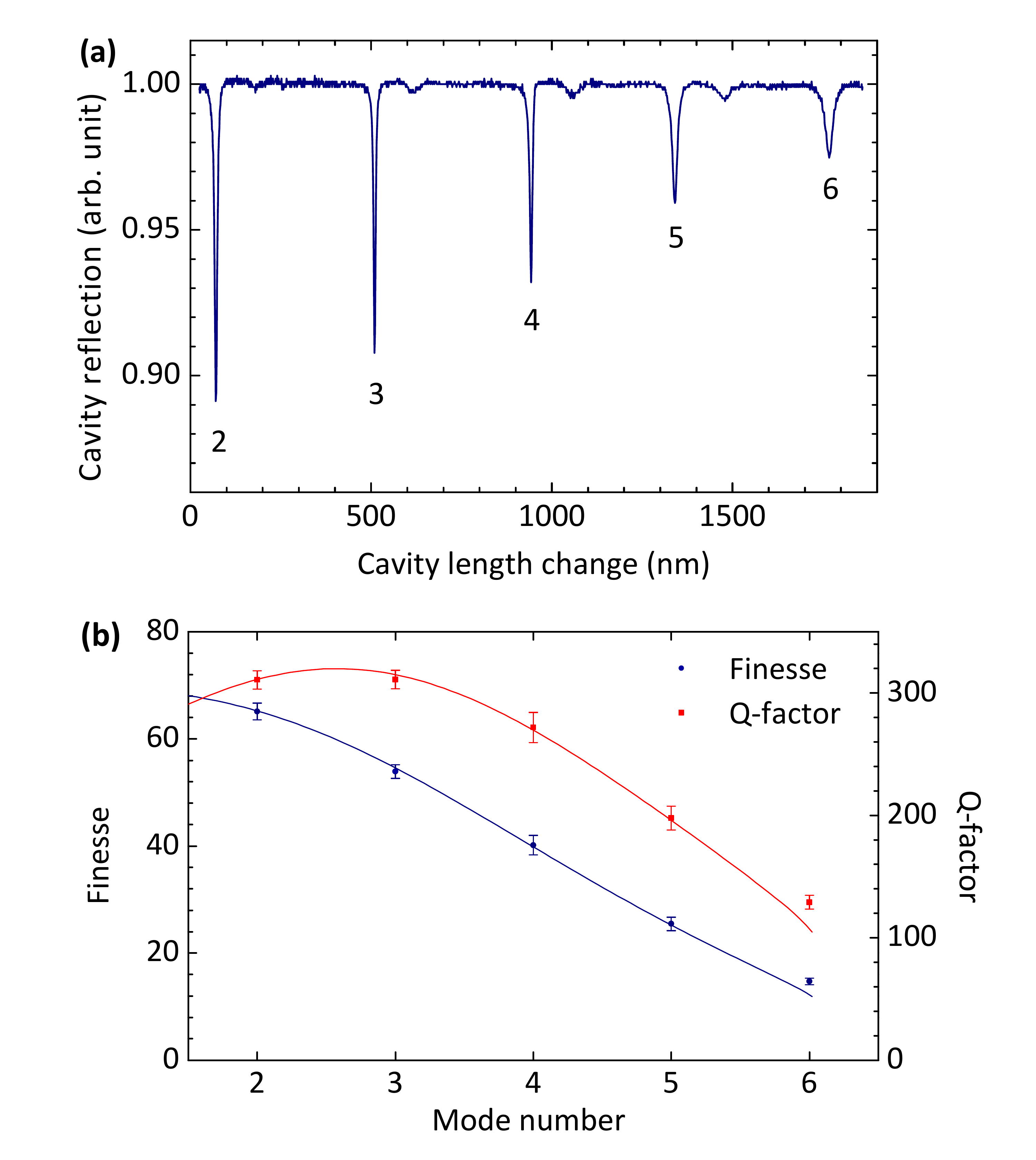}
\caption{(a) Intensity of the light reflected from the cavity as a function of the cavity length change. The small satelite resonances are attributed to higher order transverse modes. (b) Cavity $Q$ (red) and finesse $\mathcal F$ (blue). Symbols: experiment; Curves: model. }\label{linewidths}
\end{figure}

We are now prepared to analyze the cavity length spectra. In Fig.~\ref{linewidths}(a) we plot a spectrum recorded at $\lambda=785$\,nm, where the incident beam was aligned to minimize the coupling to higher transverse modes. The finesse and $Q=q_{\rm eff}\mathcal{F}$ extracted from this spectrum are shown by the symbols in Fig.~\ref{linewidths}(b). These data reveal that contrary to the common case of large cavities, $\mathcal{F}$ and $Q$ vary strongly with the mode number. 

To understand this behavior, we set up a simple model based on the propagation of a Gaussian beam between the two mirrors. We calculated the position and size of the beam waist $w_{0}$ after each round trip, whereby we took into account the loss at the finite aperture of the curved mirror (about 0.1\% of the power per round trip for the second longitudinal mode; i.e. $q$=2) simply as a scalar factor. The red curve in Fig.~\ref{linewidths}(b) shows that the $Q$ calculated from the cavity decay time is in good agreement with the experimental values. We do notice deviations for the highest mode orders, where $L$ is large enough that our model is no longer valid, and diffraction at mirror edges and losses due to beam clipping become important. The nonmonotonous behavior of $Q$ in Fig.~\ref{linewidths}(b) is the result of the competition between finite-aperture losses on the one hand and gain in the photon lifetime for larger $L$ on the other. The blue curve in Fig.~\ref{linewidths}(b) displays $\mathcal F$, which again compares well with the experimentally measured values determined from $(L_{\rm q+1}-L_{\rm q})/\delta L$. 

\subsection{Mode volume}
\label{mode volume}
The mode volume $V$ plays a central role in the performance of a cavity as suggested by the Purcell formula in the context of cavity quantum electrodynamics~\cite{Haroche-book:06}, the threshold formula in laser theory~\cite{Protsenko:99} and its role in sensing~\cite{Arnold:03}. This quantity is usually calculated as the integral of the electromagnetic energy density in the cavity mode normalized to the maximum of the field intensity. For Fabry-P\'erot resonators with perfect conductor boundaries, one can write $V=\pi w_0^2 L/4$, where $w_0$ is the Gaussian mode waist. In the presence of a DBR mirror, the extra energy leakage in the dielectric layers makes the calculation of the mode volume nontrivial, especially when $L$ becomes comparable with $\lambda$ and the DBR penetration depth~\cite{Hood:01,Greuter:14}. In order to consider this effect and the curvature of the metallic mirror, we performed full-numerical eigenmode simulations (COMSOL Multiphysics) to extract their modal properties. For computational simplicity, we neglected the thin protective layer on the micromirror, chose a simulation box size of about $4\lambda$ and imposed perfectly-matched layers and scattering boundary conditions. The typical mesh sizes were close to $\lambda/10$ near the micromirror, $\lambda/40$ in the DBR and $\lambda/4$ in the furthest regions. Figure~\ref{volume-cuts} displays cross sections of the intensity distribution for three different cavity lengths at $\lambda$=745\,nm. 

\begin{figure}[h]
\centering
\includegraphics[width=7.5cm]{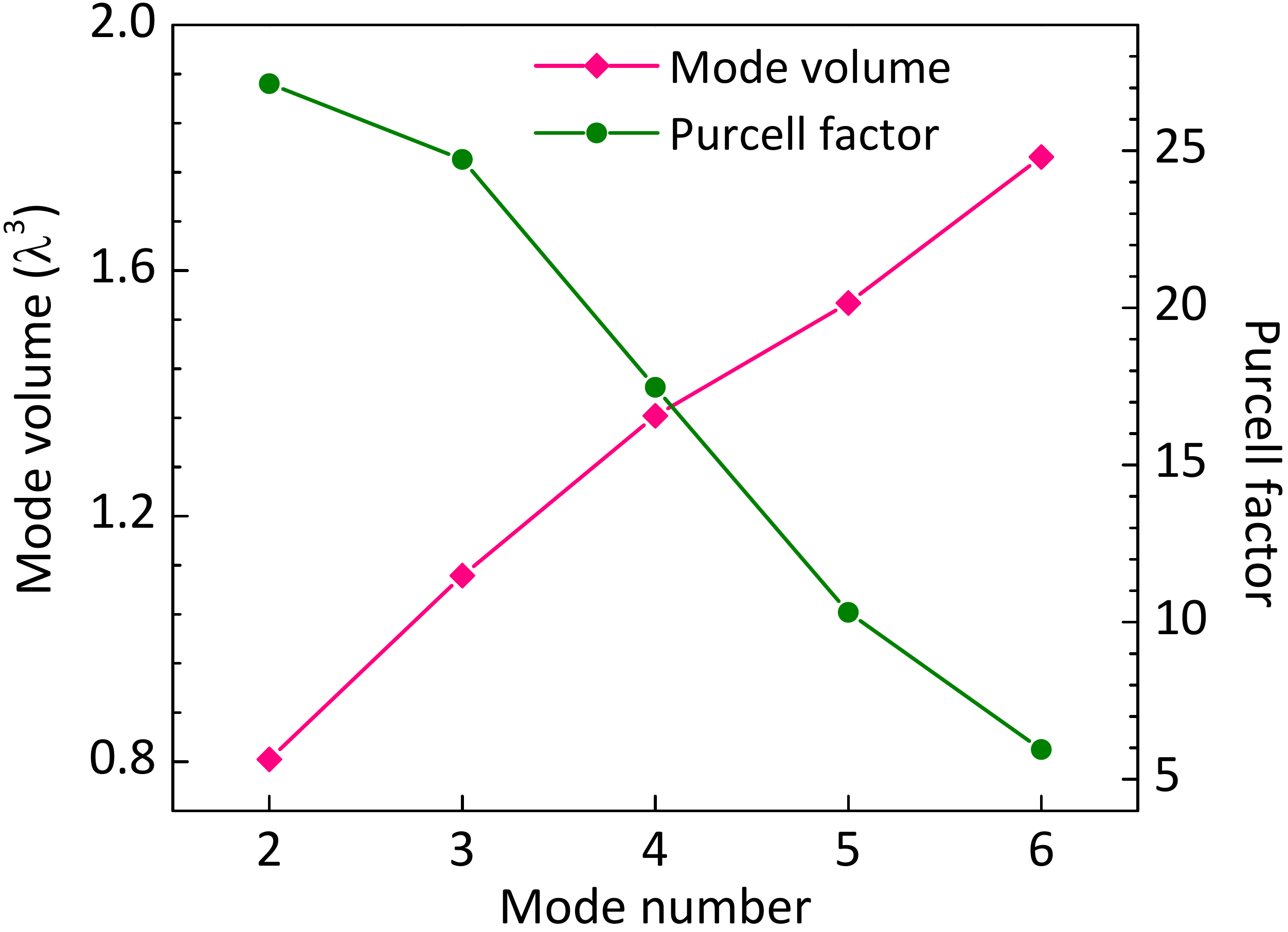}
\caption{Cavity modes volume $V$ (magenta) and Purcell factor $F_p$ (green) at $\lambda$=745\,nm.}
\label{VolumePurcell}
\end{figure}

To evaluate the mode volume from eigenmodes in the presence of radiative losses, we followed the definition provided by Ref.~\citep{Sauvan:13}, which allows for the treatment of the Purcell factor in complex environments. Figure~\ref{VolumePurcell} shows the calculated mode volume as a function of the mode number $q$ for $\lambda= 745$ nm. We find that the volume of the first accessible mode with $q=2$ is as low as $0.8 \lambda^3$, while it grows almost linearly with $q$.  

\subsection{Purcell factor and cooperativity}
The Purcell factor $F_p $ is commonly used as a measure for the ability of an optical resonator to enhance the spontaneous emission rate into a well-defined mode. An alternative formulation of the same physics is sometimes expressed as the cooperativity $C=2g^2/(\kappa\gamma)$, where $g=d\sqrt{c/(2h\epsilon \lambda V)}$ denotes the vacuum Rabi frequency of an atom with transition dipole moment $d$. Here, $\gamma$ denotes the atomic free-space radiative decay rate, and $\kappa$ signifies the cavity loss rate. After a simple manipulation, one sees that the central figure of merit remains the ratio $Q\lambda^3/V$, which can reach about 360, corresponding to $F_p \sim 27$ as shown in Fig.~\ref{VolumePurcell}. We remark that these calculations assumed a reflectivity of 94\% for the gold mirror to reach $Q$ values close to the experimental measurements. While in this work we emphasize the advantages of the low-$Q$ and low-$V$ cavity regime, higher $Q$s can be easily achieved by using enhanced metal or dielectric layers on the micromirror instead of a gold coating. Of course, one would have to keep in mind that the number of Bragg layers and their dielectric contrasts would have to be chosen properly to be compatible with the high curvature of the micromirror. We also point out that a particularly attractive application of our cavity design might be for studies at $\lambda\sim1.5\,\mu$m, where the higher reflectivity of the gold mirror coating and smaller possible values of $V/\lambda^3$ would yield higher $Q/V$.

\section{Influence of a nano-scatterer}

Having considered the basic features of our ultrasmall, broadband and tunable FPC, we now discuss experiments on coupling it to a point-like dipolar scatterer. It is known that the introduction of a foreign object in the cavity adds to the overall optical path of the photons, leading to a red shift of the cavity resonance~\cite{Arnold:03, Koenderink:05}. In the case of a subwavelength nanoparticle,  one arrives at the shift of the cavity resonance $\Delta\nu$ given by
\begin{equation}
\frac{\Delta\nu(\textbf{r})}{\nu}=-\frac{\mathcal Re (\alpha)}{2V}\frac{\vert E(\textbf{r})\vert^{2}}{\text{max}\left[\vert E(\textbf{r})\vert^{2}\right]}~,
\label{lineshifteqn}
\end{equation}
where $E(\textbf{r})$ is the electric field at position $\textbf{r}$ in the cavity \cite{Arnold:03, Koenderink:05}. Here, $\alpha$ is the complex electric polarizability of the particle, which is closely linked to its absorption and scattering cross sections \cite{Bohren-book}. In the quasi-static approximation
\begin{equation}
\alpha=\frac{\pi D^3}{2}\frac{\epsilon_{\rm p}(\lambda)-\epsilon_{\rm m}(\lambda)}{\epsilon_{\rm p}(\lambda)+2\epsilon_{\rm m}(\lambda)}~,
\label{alpha}
\end{equation} 
where $D$ is the particle diameter, and $\epsilon_{\rm p}(\lambda)$ and $\epsilon_{\rm m}(\lambda)$ are the dielectric functions of the particle and its surrounding medium, respectively. 

One might have the intuitive expectation that the introduction of a foreign object into the cavity would cause losses, thus lowering its $Q$. Although this is true in general, it has been shown that a nano-object can shift the cavity resonance without incurring a notable broadening \cite{Koenderink:05}. Indeed, recently there has been a great deal of activity to exploit the frequency shift of a high-Q cavity for sensing nanoparticles such as viruses \cite{Vollmer:08, He:11}. To investigate the effect of a nanoparticle on our cavity, we spin coated gold nanoparticles (GNP) of diameter 80 nm on the DBR with an inter-particle spacing larger than several micrometers. To ensure that we can address individual GNPs, we recorded AFM images of the DBR mirror after spin coating and identified particles that were far enough from their neighbors. By matching the AFM image of the particle distribution and its optical equivalent on our setup, we could target individual GNPs. Finally, we point out in passing that the polarizability of this GNP at the wavelength of interrogation is equivalent to a virus particle with diameter of 200 nm immersed in water. 

\subsection{Spectral shift}
Assuming that the particle is placed at the field maximum of a cavity, Eq.~(\ref{lineshifteqn}) can be written as 
\begin{equation}
\frac{\Delta\nu}{\delta\nu}=\frac{-\alpha}{2}\frac{Q}{V}~,
\label{shift-sensitivity}
\end{equation}
which expresses the ratio of a shift $\Delta\nu$ to the linewidth $\delta\nu$ of the cavity resonance profile. It is well-known that a high $Q$ facilitates the detection of a small frequency shift experienced by a narrow resonance line. In this work, we enhance the shift by using very small $V$s so that it can be detected even for broad cavity resonances. 

\begin{figure}[h]
\centering
\includegraphics[width=8.5 cm]{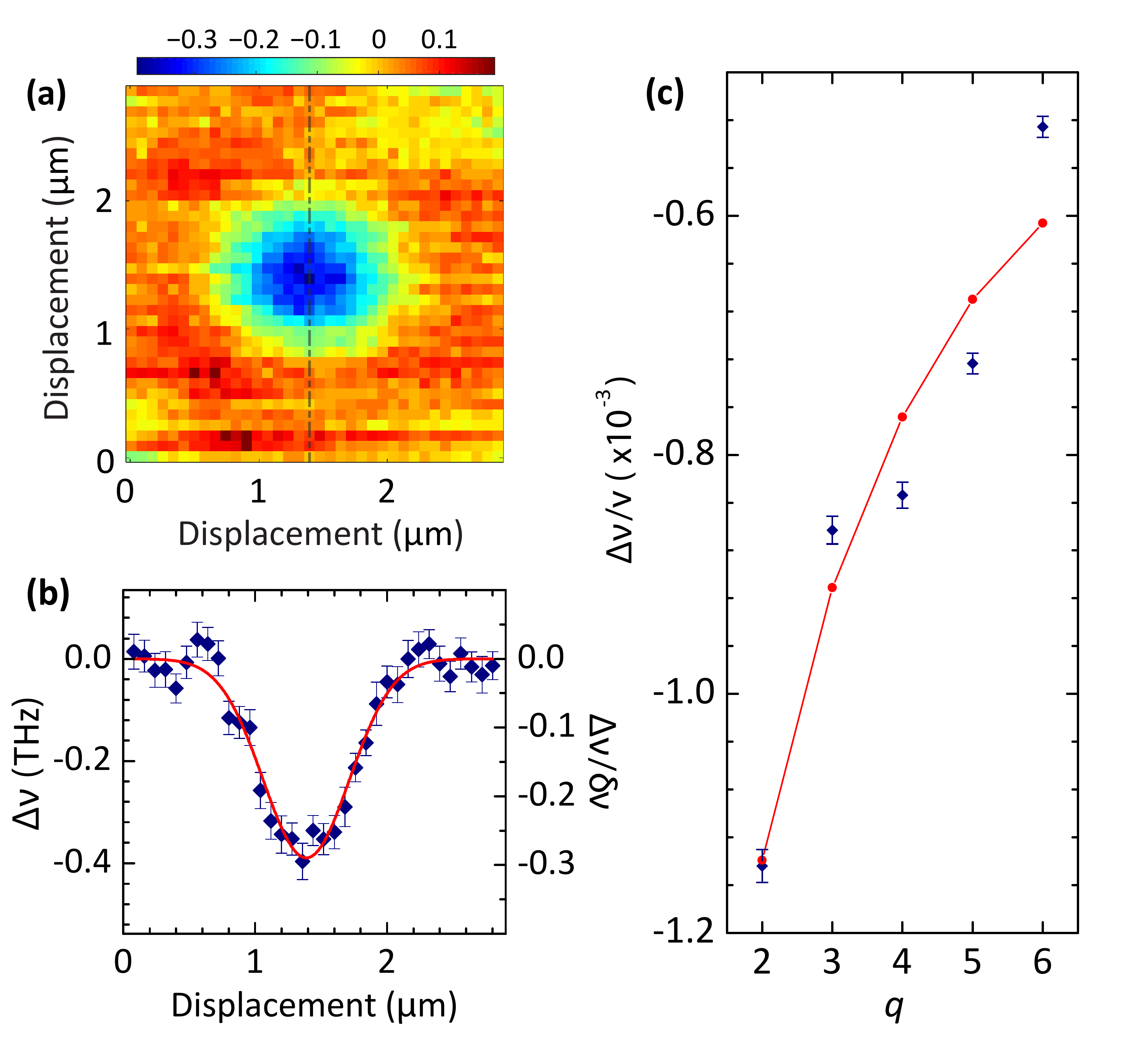}
\caption{ (a) The shift of the $q$=2 cavity resonance in THz as a function of the lateral position of a gold nanoparticle. (b) A cross section from (a), displaying FWHM $\sim 770$ nm. (c) Cavity resonance shift measured for different longitudinal modes.}\label{shift}
\end{figure}

Figure~\ref{shift}(a) shows the cavity shift as a function of the lateral position of a GNP, and Fig.~\ref{shift}(b) displays a cross section from it. We observe a shift as large as 400 GHz, equivalent to 40\% of the cavity linewidth, over a Gaussian lateral profile with $\rm FWHM \sim 770$ nm. To obtain these data, the DBR mirror holding the GNP was scanned laterally, while the cavity length was scanned at each pixel.  

The red symbols in Fig.~\ref{shift}(c) plot the maxima of cavity resonance shifts for different longitudinal modes. As expected from Eq.~(\ref{lineshifteqn}), the effect of the particle rapidly diminishes for higher $q$ modes and larger $V$s. To verify the measured data quantitatively, we have fitted them with Eq.~(\ref{lineshifteqn}), leaving $\alpha$ as a free parameter. The blue curve shows the best fit obtained for $\alpha=1.1 \times 10^6 \rm~nm^3$, which is 1.17 times larger than its expected value for a nominal GNP with a diameter of $80 \pm 6$ nm (as specified by the manufacturer), $\epsilon(\lambda)$ of gold obtained from Ref. \cite{JohnsonChristy:72}, and $\epsilon_{\rm m}=1$. However, it should be born in mind that the exact knowledge of $\alpha$ for a given GNP is highly nontrivial. First, near-field coupling to the DBR surface modifies the GNP plasmon resonance and polarizability \cite{Hakanson:08}. By using an analytical expression~\cite{pinchuk2004}, we have estimated the polarizability of the GNP to increase by 1.1 times in the presence of the DBR upper layer alone. Furthermore, variations in shape, the finite size of the particle and the resulting radiation damping effect \cite{Wokaun1982} enter beyond the simple expression of Eq.~(\ref{alpha}). Considering these effects, our experimental findings are in excellent agreement with theoretical expectations. 

\subsection{Modification of finesse}
\begin{figure}[h]
\centering
\includegraphics[width=8.5cm]{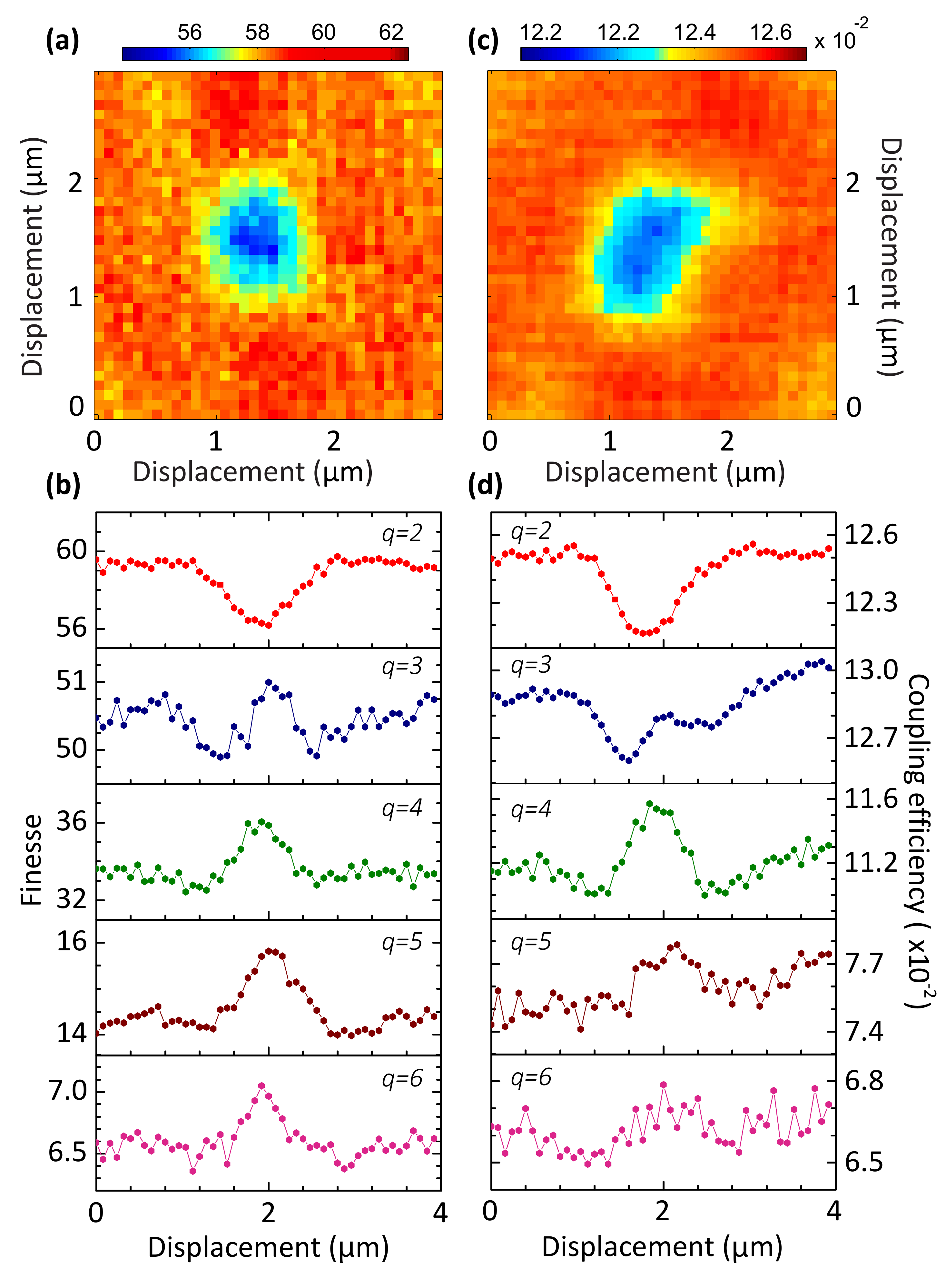}
\caption{(a) A two-dimensional map of the cavity $\mathcal F$ as a function of the particle position for $q=2$. (b) Cross sections of $\mathcal F$ vs GNP position for different cavity modes. The mode number $q$ is indicated in each measurement. (c) A two-dimensional map of the cavity coupling, obtained by normalizing the reflected power to the incident power, as a function of the particle position. (d) Cross sections of the cavity coupling efficiency for different $q$s.}\label{Q}
\end{figure}

Next, we turn to the effect of the nanoparticle on the cavity finesse for various mode orders. Since the nanoparticle scatters some of the light out of the cavity mode and has a finite absorption cross section, one can expect a degradation of $\mathcal F$. Figure~\ref{Q}(a) displays a map of $\mathcal F$ as a function of the GNP lateral position in the $q=2$ mode, and the top curve in Fig.~\ref{Q}(b) plots a cross section from it.  We find that $\mathcal F$ decreases by about 7\% from 60 to 56  in the presence of the particle while this behavior is substantially changed for higher modes (See the other plots in Fig.~\ref{Q}(b)). In fact, we find that the particle can even improve $\mathcal F$. For example, it is increased by about 10\% from 6.5 to 7 for $q=6$. 

\begin{figure}[h]
\centering
\includegraphics[width=8.5cm]{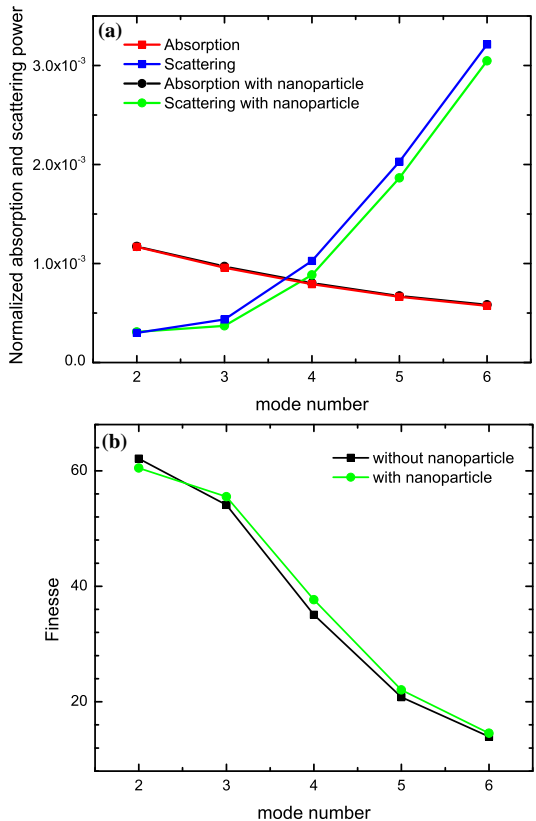}
\caption{(a) Absorption and scattering losses normalized by the cavity electromagnetic energy per optical cycle as function of the mode number. (b) Finesse in the simulations,  with a $80$ nm nanoparticle (green symbols) and without it (black).}
\label{Scatteringcoefficients_finesse}
\end{figure}

For the shortest cavities, the mirrors are so close that the light is efficiently captured after each round trip. In this case, absorption by the particle determines the quality factor and finesse. As $L$ approaches $R$, the unperturbed resonator becomes less stable and $\mathcal F$ is lowered (see Fig.~\ref{linewidths}(b)). Interestingly, in this regime the addition of a GNP ameliorates the situation. To investigate the modification of the finesse by the nanoparticle further, we have performed numerical calculations that include the nanoparticle. Here, we consider the contribution of losses caused by absorption and scattering (signified by $Q_{\rm{abs}}$ and $Q_{\rm{sca}}$) to the cavity quality factor according to $1/Q=1/Q_{\rm{abs}}+1/Q_{\rm{sca}}$. We, thus, first estimate the absorption and scattering power from simulations by calculating the overall resistive losses per optical cycle and the power flow in the directions perpendicular to the cavity axis. Figure~\ref{Scatteringcoefficients_finesse}(a) shows these quantities for the different modes normalized to the cavity energy per optical cycle. We find that without the nanoparticle, absorption losses decrease while the scattering losses grow with with $q$. A nanoparticle on the cavity axis and in contact with the DBR slightly increases the absorption (compare red and black lines), but it reduces the power scattered outside the cavity (green line) in comparison with the case without it (blue). These results confirm that the GNP acts as a mode matching antenna to improve the coupling into the cavity mode. 

To support this hypothesis further, we measured the amount of light that circulates inside the cavity by monitoring the power reflected from the DBR. Figure~\ref{Q}(c) plots the two-dimensional map of the coupling efficiency for $q=2$, and Fig.~\ref{Q}(d) displays the cross sections for various modes. The similarity of the patterns of the two data columns in Fig.~\ref{Q}(b, d) shows that the GNP strongly influences the coupling of the incoming laser beam into the cavity mode. At this point, we note that one can also perform interferometric scattering measurements (iSCAT) to detect individual GNPs without the micromirror~\cite{Piliarik:14}. Using this method, we determined a FWHM of 690 nm for the focus spot of the incident beam in the absence of the cantilever. This value agrees to within 10\% with the FWHM that we found from Fig.~\ref{shift}(b), indicating that the cavity mode waist is indeed well matched to the incoming spot size.

The black symbols in Fig.~\ref{Scatteringcoefficients_finesse}(b) display the simulated cavity finesse, whereby we adjusted the reflectivity of the gold mirror to $94\%$ to reduce the finesse by 2.4 times from its expected value for the ideal structure in order to emulate the experimental values. The trend obtained from the numerical simulations agrees very well with that presented in Fig.~\ref{linewidths}(b). In particular, we observe that the cross-over between the absorption and scattering losses at $q$=3 (see Fig.~\ref{Scatteringcoefficients_finesse}(a)) results in larger finesse in the presence of the nanoparticle (green) for higher modes, verifying the experimental observations in Fig.~\ref{Q}. In fact, even the magnitude of the change in $\mathcal{F}$ agrees well with the experimental findings.

\section{Conclusions and outlook}

The cavity regime studied in this work has brought forth different interesting phenomena, many of which had not been encountered in previous works on microcavities. These include a strong dependence of the cold cavity $Q$ and $\mathcal F$ on the longitudinal mode order $q$, the largest numerical aperture reported to date and a wavelength-sized mode waist, increase of cavity $\mathcal F$ by a nanoparticle, the combination of a sub-$\lambda^3$ mode volume and frequency tunability, and the fact that one mirror sits on a cantilever. These features are very promising for a number of future studies ranging from biophysics to quantum optics. 

One of our future goals is to modify the branching ratio of solid-state emitters such as organic molecules or color centers. The energy level scheme of such systems usually involves a narrow transition on the so-called zero-phonon line (ZPL) accompanied by a broad phonon wing and transitions among other vibrational levels. In order to have a strong transition for quantum optics experiments, it is desirable to exploit the Purcell effect to enhance the ZPL and thus improve its branching ratio with respect to the other transitions. In our current measurements, $(\lambda^3Q/V)$ reaches a value of $\sim 375$, corresponding to $F_{\rm p}=29$, which would be sufficient for turning a typical branching ratio of about 30\% for the ZPL of aromatic molecules to about 92\%. Such a nearly perfect two-level system can then act as an efficient source of single narrow-band photons~\cite{Lettow:07,Trebbia:09}. 

Another potential use of our cavity concept is for the achievement of few-photon nonlinear effects on single quantum emitters~\cite{Chang:14}. Examples of such an effect are three-photon mixing, stimulated Rayleigh scattering and hyper-Raman scattering~\cite{Boyd-book}. While there have been a few reports of direct cavity-free optical nonlinear studies on single molecules and quantum dots~\cite{Lounis:97, Xu:07, Maser:15}, enhancements of the optical field in the cavity and of the radiation of a quantum emitter into its mode can boost these effects further. Here, a large cavity bandwidth is of crucial importance since simultaneous coupling to several wavelengths would be otherwise not possible. 

The broad resonance of a microcavity is also a great asset for studying cooperative effect among many quantum emitters with slightly different resonance frequencies or with large homogeneous lines as it occurs in solid-state systems. An example of such phenomena concerns Rabi splittings and coupling to polaritonic states at room temperature~\cite{Lidzey:00, Kena-Cohen:08, Schwartz:11}. Although the coupling of each individual molecule or exciton is negligible in these systems, strong Rabi splitting are observed when a large number of emitters join efforts. The combination of a large cooperativity, small $V$ and low $Q$ would allow one to examine polaritonic physics with far fewer emitters than previously accessible both at room temperature and at cryogenic conditions. 

A further practical advantage of a considerable cooperativity factor combined with a low $Q$ is that the experimental setup becomes much less sensitive to mechanical instabilities. This is particularly helpful for experiments in high vacuum or in cryostats, where vibrations are omnipresent. Indeed, the arrangement of the cavity presented here could be highly interesting for use in atom~\cite{Trupke:07} and nanoparticle trapping~\cite{Hu:10}, especially for on-chip schemes~\cite{Colombe:07}. For example, the antinode in Fig.~\ref{volume-cuts}(a) is ideally suited for this purpose, keeping the particle sufficiently far from mirror surfaces to avoid van der Waals forces~\cite{Hinds:91}. 

The cantilever-based nature of our experimental arrangement also holds great promise in the context of optomechanics. Let us consider a silicon cantilever of thickness 0.7 $\mu \rm m$, length 10 $\mu \rm m$ and width 5 $\mu \rm m$, yielding a mechanical oscillation frequency of about 5 MHz and mass of $10^{-13}$~kg. A micromirror on such a cantilever forming a cavity length $L=1~\mu \rm m$ would result in a nearly maximal field per photon $\sqrt{\hbar \omega/2\epsilon_0V}$ and an optomechanical coupling strength between the photon and a single phonon of $g^{\rm om}_0/2\pi > 10^6$~Hz, which is comparable with the best reported values \cite{Aspelmeyer:14,Chan:11}. For these experiments, a higher Q would further improve the cavity performance. As mentioned earlier, these can be readily increased by using dielectric coatings instead of gold. To respect the small radius of curvature of the micromirror, one would use materials with strong dielectric contrasts to reduce the number of the necessary Bragg layers. 

Other applications of the work presented here are in optical sensing and analytics. Over the past decades, a large body of literature has been devoted to the use of optical micro- and nanostructures such as waveguides, cavities and plasmonic nanoantennas for detecting small traces of biological matter~\cite{Vollmer:12}. The existing works on microcavity sensing use fairly large mode volumes and require, therefore, high $Q$s to reach sufficient sensitivity. Recently, we demonstrated that single small proteins can be detected via their direct scattering in a label-free fashion if only diffraction-limited imaging is implemented~\cite{Piliarik:14}. In a nutshell, we have shown that the scattering cross section of a protein, as small as it is, is large enough to extinguish a measurable amount of power from a laser beam. If one now maintains a tight focus and circulates the laser beam in a cavity to allow for repeated interactions with the analyte species, one directly wins in the detection sensitivity by $\cal F$ folds. Based on a similar argument, our tunable and scannable microcavity can also find applications in Raman microscopy, where small cross sections would be compensated by larger excitation intensities. Again, a large cavity bandwidth is essential to allow a simultaneous coupling of the excitation beam and the Raman signal.

The physics of microlasers is another area that would benefit from our cavity design. Since its invention more than half a century ago, there has been a continuous stream of reports on various fundamental aspects of laser physics, for example of lasing with minimal gain or with very small thresholds. It turns out that again an important figure of merit is given by $Q/V$~\cite{Svelto-book}. Here, a higher $Q$ increases the degree of coherence although it also minimizes the overlap between the homogeneous spectrum of the gain medium and the cavity mode, resulting in higher thresholds. A lower $V$ (and thus a higher finesse) pushes away the modes to reduce mode competition. In addition, a large Purcell factor ensures a high $\beta$, which determines the fraction of the overall spontaneous emission funneled in the mode of interest according to $\beta=F_{\rm p}/(1+F_{\rm p})$. Therefore, the regime of our microcavity, simultaneously keeping a low $Q$, low $V$ and high $F_{\rm p}$, has promise in the development of new microlasers, especially in combination with fluidics~\cite{psaltis_developing_2006}. Aside from their academic interest, low-threshold microlasers are also highly in demand for applications, where miniaturization and scaling play a role. 

We acknowledge financial support from the European Union (ERC Advanced Grant SINGLEION and project UnivSEM of the seventh framework program under Grant Agreement No. 280566), Alexander von Humboldt Foundation (Humboldt Professur) and the Max Planck Society. We thank Harald Haakh, Ioannis Chremmos and Florian Marquardt for fruitful discussions and Maksim Medvedev, Lothar Meier and Marek Piliarik for technical support at various stages of this project. We also thank Sara Mouradian for her contribution to the very early stage of this work. H.K. and D.W. contributed equally to this work.


\begin{thebibliography}{57}
\bibitem[{foo()}]{footnote-Kelkar}
\bibinfo{note}{We follow the convention that $\phi<0$ so that the reflection
  surface is inside the mirror. The reflection phase lies in the second
  quadrant for the gold mirror. This convention is different from that of
  Ref.~\citep{Bennett:64}.}
\end{thebibliography}
\end{document}